\documentclass[sigconf]{acmart}

\AtBeginDocument{%
  }

\usepackage{tikz}
\usetikzlibrary{shapes,arrows,positioning,calc,fit,backgrounds}
\usepackage{algorithm}
\usepackage{algpseudocode}
\usepackage{booktabs}
\usepackage{multirow}
\usepackage{xcolor}
\usepackage{pifont}
\usepackage{graphicx}
\graphicspath{{figures/}}
\newcommand{\cmark}{\ding{51}}
\newcommand{\xmark}{\ding{55}}

\newcommand{\sys}{SysOM-AI}
\begin{document}

\title{\sys{}: Continuous Cross-Layer Performance Diagnosis for Production AI Training}

\renewcommand{\authorsaddresses}{}  
\author{Yusheng Zheng$^{*,2}$, Wenan Mao$^{*,1}$, Shuyi Cheng$^1$, Fuqiu Feng$^1$, Guangshui Li$^1$, Zhaoyan Liao$^1$, Yongzhuo Huang$^1$, Zhenwei Xiao$^1$, Yuqing Li$^1$, Andi Quinn$^{\dagger,2}$, Tao Ma$^{\dagger,1}$}
\affiliation{\institution{$^1$Alibaba Group}\country{China}}
\affiliation{\institution{$^2$UC Santa Cruz}\country{USA}}

\renewcommand{\shortauthors}{Zheng, Mao, et al.}
\renewcommand\footnotetextcopyrightpermission[1]{}

\begin{abstract}
Performance diagnosis in production-scale AI training is challenging because subtle OS-level issues can trigger cascading GPU delays and network slowdowns, degrading training efficiency across thousands of GPUs.
Existing profiling tools are limited to single system layers, incur prohibitive overhead (10--30\%), or lack continuous deployment capabilities, resulting in manual analyses spanning days.
We argue that continuous, cross-layer observability enabled by OS-level instrumentation and layered differential diagnosis is necessary to address this gap.
We introduce \sys{}, a production observability system that continuously integrates CPU stack profiling, GPU kernel tracing, and NCCL event instrumentation via adaptive hybrid stack unwinding and eBPF-based tracing, incurring less than 0.4\% overhead.
Deployed at Alibaba across over 80,000 GPUs for more than one year, \sys{} helped diagnose 94 confirmed production issues, reducing median diagnosis time from days to approximately 10 minutes.
\end{abstract}



\maketitle
\let\thefootnote\relax\footnotetext{$^*$Equal contribution (co-first authors). $^\dagger$Corresponding authors: boyu.mt@taobao.com, aquinn1@ucsc.edu.}

\section{Introduction}
\label{sec:intro}

Large-scale AI training jobs increasingly dominate modern production datacenters, often consuming thousands of GPUs simultaneously.
At such scale, even minor performance inefficiencies caused by OS-level scheduling, GPU thermal behavior, or network contention can cascade into substantial slowdowns, impacting overall training throughput~\cite{patel2024revisiting}.
Prior characterization studies report that such issues are a major source of training job inefficiencies, with diagnosis often taking days~\cite{hu2024characterization}.
Diagnosing and mitigating these issues remains a significant operational challenge.

Current performance profiling tools predominantly focus on single-layer visibility (Table~\ref{tab:comparison}): GPU-specific profilers (e.g., NVIDIA Nsight~\cite{nsight}) monitor GPU execution but impose 10--30\% overhead~\cite{nsight_overhead}; CPU-focused tools (e.g., Strobelight~\cite{strobelight}, async-profiler~\cite{asyncprofiler}) efficiently capture call stacks but lack visibility into GPU or distributed NCCL events; and framework-level profilers (e.g., PyTorch Profiler~\cite{pytorchprofiler}) ignore critical OS-level behaviors.
These tools cannot continuously run in large-scale production due to prohibitive overhead or incomplete system coverage.
Consequently, operators must manually correlate fragmented data from multiple sources, significantly prolonging diagnosis.

We argue that continuous, cross-layer observability (capturing and correlating CPU, GPU, and network activity simultaneously) is essential to identify root causes of performance degradation in production AI training.
Building on this insight, we develop \sys{}, a production profiling system that leverages OS-level eBPF instrumentation to enable lightweight, always-on collection of CPU stacks (including full kernel call chains), GPU kernel timings, and NCCL collective events.
A central analysis service correlates these signals across all ranks and performs \emph{layered differential diagnosis}: comparing GPU, CPU, and OS profiles between anomalous and healthy ranks to systematically isolate the root-cause layer, or comparing against historical baselines to catch uniform degradation.
Achieving production-scale deployment required addressing several challenges, including minimizing profiling overhead via adaptive hybrid stack unwinding, implementing centralized symbol resolution to avoid per-node memory pressure, and developing framework-agnostic NCCL tracing without debug symbols.

\sys{} has been deployed for over one year across Alibaba's production fleet encompassing more than 80,000 GPUs and 10,000+ nodes, processing $\sim$400\,TiB of profiling data daily.
In the six-month evaluation period reported in this paper, it helped diagnose 94 confirmed performance issues (GPU thermal throttling, NIC soft-interrupt contention, VFS lock contention, logging overhead) that were not detected by existing single-layer tools, reducing median diagnosis time from days to approximately 10 minutes.

\begin{table}[t]
\centering
\caption{Comparison with AI profiling tools. \cmark = supported, \xmark = not supported. ``Kernel'' = OS kernel-level signals. ``Scope'' = single process vs.\ cluster-wide.}
\label{tab:comparison}
\small
\begin{tabular}{@{}lcccccc@{}}
\toprule
\textbf{Tool} & \textbf{Py} & \textbf{C++} & \textbf{Kern.} & \textbf{Scope} & \textbf{GPU} & \textbf{Ovhd.} \\
\midrule
MegaScan~\cite{megascan}       & \cmark & \xmark & \xmark & Proc & \xmark & Low \\
MegaScale~\cite{jiang2024megascale}      & \xmark & \xmark & \xmark & Proc & \cmark & Low \\
DeepSpeed Prof.~\cite{deepspeed}  & \cmark & \xmark & \xmark & Proc & \cmark & Med \\
Nsight Sys.~\cite{nsight}    & \xmark & \cmark & \xmark & Proc & \cmark & High \\
Strobelight~\cite{strobelight}    & \xmark & \cmark & \xmark & Proc & \xmark & Low \\
async-profiler~\cite{asyncprofiler} & \xmark & \cmark & \xmark & Proc & \xmark & Low \\
\midrule
\textbf{\sys{}}          & \cmark & \cmark & \cmark & Cluster & \cmark & Low \\
\bottomrule
\end{tabular}
\par\smallskip
\raggedright\scriptsize
Among these tools, none collect kernel-level OS signals such as interrupt handling or scheduler activity.
\sys{} adds continuous cluster-scoped observability with OS-kernel visibility.
\end{table}

This paper makes the following contributions:
\begin{enumerate}
\item We design and implement \sys{}, a continuous cross-layer performance profiling system for production AI training, integrating CPU, GPU, and NCCL observability at less than 0.4\% overhead (\S\ref{sec:design}--\ref{sec:implementation}).

\item We propose a layered differential analysis methodology that systematically isolates root causes by comparing across ranks (for stragglers) and against historical baselines (for uniform degradation), using a per-communication-group \emph{CPU waterline} (\S\ref{sec:differential-diagnosis}).

\item We implement adaptive hybrid stack unwinding, centralized symbol resolution, and framework-agnostic NCCL tracing via eBPF, enabling always-on profiling with low overhead (\S\ref{sec:hybrid-unwinding}--\ref{sec:nccl-observability}).

\item We report on a one-year production deployment across 80,000+ GPUs, presenting end-to-end case studies demonstrating cross-layer diagnosis of issues invisible to existing tools (\S\ref{sec:evaluation}).
\end{enumerate}

\section{Background and Motivation}
\label{sec:background}

We first describe the performance characteristics of distributed AI training, then use a production incident to motivate the design of \sys{}.

\subsection{AI Training Performance Characteristics}

Modern large-scale AI training uses synchronous data or pipeline parallelism, where each iteration comprises forward pass, backward pass, gradient reduction via collectives (e.g., AllReduce, ReduceScatter), and parameter update~\cite{jiang2024megascale}.
Because collectives are blocking, the slowest rank in a communication group determines the iteration time for all ranks.
Even small per-rank delays compound at scale: a 0.3\,ms slowdown per iteration on a single rank wastes over 70 GPU-hours per day in a 1,000-GPU job running at 100\,ms/iteration.
Prior work has shown that such performance issues are common and stem from diverse root causes spanning GPU hardware, OS kernel, network, and application runtime~\cite{patel2024revisiting, hu2024characterization}, making it difficult to determine \emph{which} layer is responsible without profiling data from all layers simultaneously.

\subsection{Motivating Example}
\label{sec:motivating-example}

Consider a scenario from our production fleet: rank~4 in an 8-rank communication group enters each NCCL collective 0.6\,ms late, slowing the entire group.
An operator's first instinct is to check GPU utilization, but \texttt{nvidia-smi} reports 100\% on all ranks, and GPU kernel times match within 1\%.
Network monitoring shows no packet loss or latency anomaly.
A CPU profiler (\texttt{perf}) is attached, but because \texttt{asm\_common\_interrupt} is compiled without frame pointers, FP-based unwinding truncates the stack at the interrupt boundary, hiding everything below it.
The operator sees elevated CPU time in an unrelated function due to misattribution, and spends two days investigating a false lead.

The actual root cause was NIC soft-interrupts (\texttt{NET\_RX\_SOFTIRQ}) bound to a CPU core shared with the NCCL communication thread.
Diagnosing this required three capabilities that no existing tool provides together:
(1)~\emph{cross-rank comparison} to see that GPU profiles match but CPU profiles diverge,
(2)~\emph{accurate kernel stacks} to reveal the full interrupt chain from \texttt{asm\_common\_interrupt} through \texttt{do\_softirq} to \texttt{napi\_gro\_receive}, and
(3)~\emph{always-on collection} because the issue was intermittent and could not be reproduced with a profiler attached after the fact.
We detail this case fully in \S\ref{sec:eval-cases} (Case~2).

\subsection{Challenges}
\label{sec:challenges}

This example illustrates four fundamental gaps in existing tooling, each of which \sys{} addresses with a dedicated mechanism.

\textbf{C1: No cross-layer diagnosis methodology.}
Existing tools operate on individual layers: Nsight Systems~\cite{nsight} profiles GPU kernels, Strobelight~\cite{strobelight} profiles CPU stacks, and network counters track link-level statistics.
No tool systematically compares profiles across GPU, CPU, and OS layers to isolate which layer is responsible for a slowdown.
Operators must manually correlate fragmented data, resulting in diagnosis times typically spanning days.
\sys{} addresses this with layered differential diagnosis (\S\ref{sec:differential-diagnosis}).

\textbf{C2: Framework-coupled NCCL instrumentation.}
Cross-rank diagnosis requires per-collective timing to detect stragglers.
Prior tools such as MegaScan~\cite{megascan} obtain this by instrumenting NCCL at the framework level, tightly coupling to Megatron's code paths; switching frameworks requires re-instrumentation.
Furthermore, production NCCL builds ship without debug symbols, making it difficult to identify communication groups and match collective instances.
\sys{} addresses this with framework-agnostic eBPF-based NCCL tracing (\S\ref{sec:nccl-observability}).

\textbf{C3: Inaccurate stacks at production-acceptable overhead.}
Even within a single layer, CPU profilers produce unreliable data.
FP-based unwinding achieves low frame accuracy on production AI workloads because functions compiled with \texttt{-fomit-frame-pointer} (the GCC/Clang default at \texttt{-O2}) cause truncated or misattributed stacks; in one production case, this led to 50\% of CPU samples being misattributed to a single function (\S\ref{sec:eval-accuracy}).
DWARF-based unwinding is accurate but significantly more expensive per sample, consuming several percent of a core at 100\,Hz, which is unacceptable for always-on use.
\sys{} addresses this with adaptive hybrid FP+DWARF unwinding (\S\ref{sec:hybrid-unwinding}).

\textbf{C4: Symbol resolution breaks at scale.}
Production binaries are typically stripped, with debug symbols stored separately.
Loading full symbol tables (600\,MB--1\,GB for large libraries) on every profiled node causes out-of-memory failures, and node-side nearest-address matching produces incorrect attributions when symbol tables are sparse (\S\ref{sec:eval-accuracy}).
\sys{} addresses this with centralized Build-ID-indexed symbol resolution (\S\ref{sec:symbol-resolution}).

\section{System Design}
\label{sec:design}

\begin{figure}[t]
\centering
\includegraphics[width=\columnwidth]{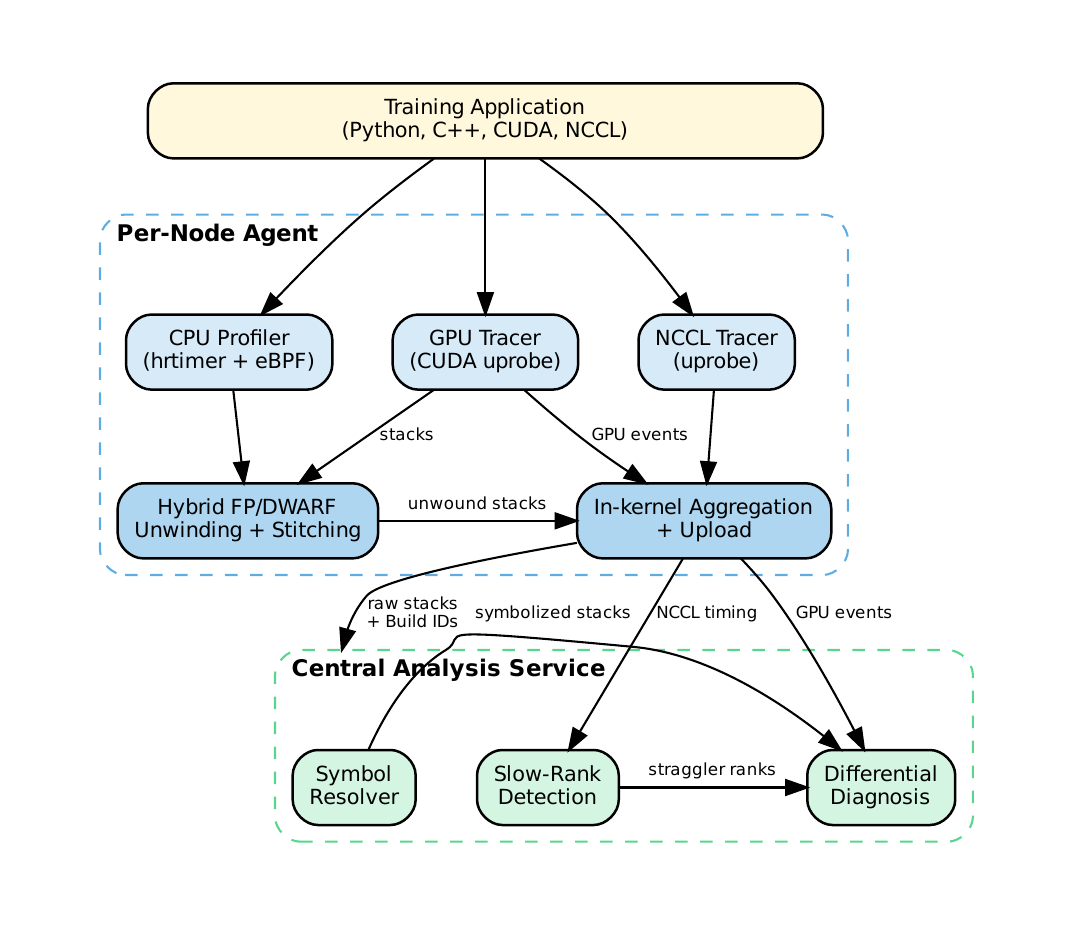}
\caption{\sys{} architecture. Each node runs an agent that collects CPU stacks (hrtimer + eBPF), GPU kernel timings (CUDA uprobe), and NCCL collective events (uprobe). Stacks are unwound via adaptive hybrid FP/DWARF, stitched across Python and C++ frames, aggregated in-kernel, and uploaded to a central service for symbol resolution, slow-rank detection, and layered differential diagnosis.}
\label{fig:architecture}
\end{figure}

To address the challenges identified in \S\ref{sec:challenges}, \sys{} combines always-on eBPF-based data collection with a centralized analysis pipeline.
Figure~\ref{fig:architecture} shows the architecture.


At a high level, \sys{} operates in three stages: (1)~continuous data collection via eBPF-based probes on every node, (2)~centralized symbol resolution and aggregation, and (3)~layered differential diagnosis that pinpoints root causes across system layers.
We first describe the diagnosis methodology (\S\ref{sec:differential-diagnosis}), then the three enabling mechanisms: NCCL observability (\S\ref{sec:nccl-observability}), adaptive stack unwinding (\S\ref{sec:hybrid-unwinding}), and centralized symbol resolution (\S\ref{sec:symbol-resolution}).

\subsection{Layered Differential Diagnosis}
\label{sec:differential-diagnosis}

\textbf{CPU waterline.}
\label{sec:cpu-waterline}
We define the \textbf{CPU waterline} as the per-communication-group baseline CPU profile computed over a sliding window of the most recent $W$ iterations (default $W=100$).
For each function $f$ in communication group $g$, we compute the mean CPU fraction $\mu_f^g$ and standard deviation $\sigma_f^g$ across all ranks in $g$.
A rank $r$ is flagged when any function exceeds $\mu_f^g + k\sigma_f^g$ (default $k=2$).
The waterline is computed over \emph{all} ranks in the group simultaneously; no prior partitioning into ``healthy'' and ``unhealthy'' is needed.
Stragglers are statistical outliers: a single anomalous rank among $N$ ranks shifts $\mu$ by only $1/N$ and increases $\sigma$ modestly, so the outlier's functions still exceed the threshold.
For $N \geq 8$ (typical communication group size), the influence of one straggler on the group statistics is bounded (\S\ref{sec:eval-cases}).

\textbf{Slow-rank detection.}
Using the NCCL instrumentation described in \S\ref{sec:nccl-observability}, for each collective \sys{} records the host-side entry timestamp and the GPU-side duration.
Cross-rank clock alignment uses the collective's barrier semantics: since all ranks must enter and exit, the earliest entry and latest exit provide natural reference points for aligning cross-rank timestamps.
A rank is flagged as a straggler if its entry time exceeds $\mu + k\sigma$ (default $k=2$) over a sliding window of $W$ iterations (\S\ref{sec:eval-cases}).

\textbf{Cross-rank differential diagnosis.}
When a straggler is detected, \sys{} generates layer-by-layer differential profiles:
\begin{sloppypar}
\begin{enumerate}
    \item \textbf{GPU diff}: Compare GPU kernel execution times. Uniform slowdown across all kernels suggests hardware issues (thermal throttling, memory errors). Slowdown in specific kernels suggests software issues (operator-level change).
    \item \textbf{CPU diff}: If GPU profiles match, compare CPU flame graphs. New hot functions or increased time in specific paths reveal host-side interference (interrupts, scheduling, I/O).
    \item \textbf{OS diff}: If application-level CPU profiles match, compare OS subsystem signals: interrupt counts (\texttt{/proc/interrupts}), scheduler latency (\texttt{sched\_stat} tracepoints), and NUMA migration events, which may not appear in sampled flame graphs due to their brief, high-frequency nature.
\end{enumerate}
\end{sloppypar}

\textbf{Temporal baseline comparison.}
When no straggler is detected but absolute iteration time increases, \sys{} compares the current per-group flame graph against a historical baseline stored in the centralized log service.
Functions whose CPU fraction increased by more than $\delta$ (default 0.5\%) relative to the baseline are flagged as degradation candidates.
This complements cross-rank analysis: cross-rank comparison identifies \emph{which rank} is anomalous, while temporal comparison identifies \emph{when} performance changed and \emph{what code path} is responsible.

\subsection{Framework-Agnostic NCCL Observability}
\label{sec:nccl-observability}

The diagnosis pipeline above relies on per-collective timing data.
Obtaining this data in production requires addressing three problems.

\textbf{Framework-agnostic tracing.}
Rather than instrumenting NCCL at the framework level (\S\ref{sec:challenges}, C2), \sys{} intercepts NCCL calls at the library boundary, achieving framework-agnostic tracing that works identically across Megatron, DeepSpeed, ms-swift, and other frameworks.

\textbf{Group identification without debug symbols.}
Communication group identification typically requires parsing \texttt{ncclComm} structures using debug information, which production NCCL builds lack (\S\ref{sec:challenges}, C2).
\sys{} pre-parses NCCL structure layouts at known version-specific offsets (currently NCCL~2.14--2.21 and ACCL), enabling group identification without runtime debug information at the cost of a configuration update when NCCL's internal layout changes.

\textbf{Collective-instance separation.}
Within a communication group, multiple collectives may be in flight simultaneously.
The standard approach uses the \texttt{opCount} field in \texttt{ncclComm}, but for point-to-point operations this counter resides in GPU memory, making it expensive to read.
\sys{} instead exploits \emph{temporal overlap}: operations that overlap in time across ranks are identified as belonging to the same collective instance, avoiding GPU memory access.

\subsection{Adaptive Hybrid FP+DWARF Stack Unwinding}
\label{sec:hybrid-unwinding}

Accurate CPU flame graphs are central to the differential diagnosis in \S\ref{sec:differential-diagnosis}.
The key insight is that FP-based unwinding is correct for the majority of functions in typical binaries, i.e., those compiled with \texttt{-fno-omit-frame-pointer} or that otherwise preserve the frame pointer convention.
Approximately 20\% of functions in our production binaries require DWARF, primarily C++ libraries compiled with optimizations that omit frame pointers.
\sys{} learns per-function which method is needed and caches the decision, amortizing DWARF's cost over time.

\textbf{Unwinding method markers.}
Each function is identified by its Build ID and code offset.
\sys{} maintains a hash map \texttt{Map<(BuildID, Offset) $\rightarrow$ Marker>} where \texttt{Marker} $\in$ \{\texttt{unmarked}, \texttt{fp}, \texttt{dwarf}\}.
Algorithm~\ref{alg:hybrid} describes the unwinding procedure.

\begin{algorithm}[t]
\caption{Adaptive Hybrid Stack Unwinding}
\label{alg:hybrid}
\small
\begin{algorithmic}[1]
\Require Registers: PC, SP, FP
\Ensure Call stack $S$
\State $S \gets []$
\While{PC is in a mapped executable region}
    \State $m \gets \text{GetMarker}(\text{BuildID}(\text{PC}), \text{Offset}(\text{PC}))$
    \If{$m = \texttt{unmarked}$}
        \State $(pc', sp', fp') \gets \text{UnwindFP}(\text{PC}, \text{SP}, \text{FP})$
        \If{\Call{ValidateCallerPC}{$pc'$, $sp'$, PC}}
            \State $\text{SetMarker}(\text{PC}, \texttt{fp})$
        \Else
            \State $(pc', sp', fp') \gets \text{UnwindDWARF}(\text{PC}, \text{SP})$
            \State $\text{SetMarker}(\text{PC}, \texttt{dwarf})$
        \EndIf
    \ElsIf{$m = \texttt{fp}$}
        \State $(pc', sp', fp') \gets \text{UnwindFP}(\text{PC}, \text{SP}, \text{FP})$
    \Else \Comment{$m = \texttt{dwarf}$}
        \State $(pc', sp', fp') \gets \text{UnwindDWARF}(\text{PC}, \text{SP})$
    \EndIf
    \State $S.\text{append}(pc')$
    \State $(\text{PC}, \text{SP}, \text{FP}) \gets (pc', sp', fp')$
\EndWhile
\end{algorithmic}
\end{algorithm}

\textbf{Validation.} The \textsc{ValidateCallerPC} function checks whether the unwound return address $pc'$ falls within a mapped executable region by consulting the process's memory map (\texttt{/proc/\allowbreak[pid]/\allowbreak maps}).
Specifically, it verifies two conditions: (1)~$pc'$ is within the \texttt{[start,~end)} range of some executable ELF segment, and (2)~the stack pointer $sp'$ is monotonically increasing (i.e., the stack is being unwound upward).
If either check fails, the FP result is invalid, typically because the function was compiled with \texttt{-fomit-frame-pointer}, causing the frame pointer register to hold a general-purpose value rather than the caller's frame address.

\textbf{Marker convergence and stability.}
In our production workloads, the majority of distinct call-site markers converge within the first profiling window.
Functions encountered only in rare code paths (error handling, initialization) fall back to DWARF on first encounter and are cached thereafter.

Markers are \emph{stable} because a function's frame-pointer behavior is determined at compile time and does not change at runtime.
The two exceptions are dynamically loaded libraries (\texttt{dlopen}) and JIT-compiled code, whose functions are initially unmarked and go through the same convergence process; we describe the detection mechanisms in \S\ref{sec:implementation}.

\textbf{Cost analysis.}
\label{sec:cost-model}
Since the majority of functions use the cached FP path and only a small fraction require DWARF, the steady-state per-sample cost is close to pure FP unwinding.
DWARF frames tend to occur higher in the stack and contribute fewer frames per sample, further reducing the overhead in practice.
We quantify this in \S\ref{sec:eval-overhead}.

\subsection{Centralized Deferred Symbol Resolution}
\label{sec:symbol-resolution}

Even with accurate unwinding, raw address stacks must be symbolized to produce readable flame graphs.

\textbf{Problem.}
Production AI binaries (e.g., the Pangu distributed storage client, PyTorch shared libraries) are typically stripped, with debug symbols available only in separate files.
Loading full symbol tables (600\,MB--1\,GB for large internal libraries) on every profiled node causes OOM when multiple processes are profiled simultaneously.
Furthermore, node-side ``nearest lower address'' symbol matching produces incorrect attributions when symbol tables are sparse; we document a concrete instance in \S\ref{sec:eval-accuracy} where this caused 50\% of CPU samples to be misattributed.

\textbf{Design.}
The node agent collects raw address stacks along with the Build ID of each loaded binary.
At upload time, the agent checks the central symbol repository for an existing symbol file keyed by Build~ID; if absent, it extracts and uploads the debug symbols.
The central resolver performs symbol lookup using the complete symbol table, producing accurate function names via $O(\log n)$ lookup without loading the entire file into memory.
In our deployment, the repository stores over 170,000 distinct Build IDs in a single region.

\section{Implementation}
\label{sec:implementation}

The node agent is implemented in C (eBPF programs) and Rust (userspace daemon); the central service in Go.

\textbf{eBPF programs and agent communication.}
The agent uses three eBPF program types:
(1)~\texttt{perf\_event} attached to \texttt{hrtimer} for periodic CPU sampling (configurable 10--999\,Hz; default 99\,Hz to avoid lock-step aliasing with timer interrupts),
(2)~\texttt{uprobe} on CUDA runtime functions (\texttt{cuLaunch\-KernelEx}) for GPU kernel launch interception, and
(3)~\texttt{uprobe} on NCCL entry points (\texttt{nccl\-AllReduce}, \texttt{nccl\-Reduce\-Scatter}, etc.) for collective tracking.
For in-kernel aggregation, the eBPF program hashes each stack trace and increments a per-stack counter in a BPF hash map. The userspace daemon drains the map every 5\,s, reducing the data volume by 10--50$\times$ compared to per-sample streaming.
The node agent communicates with training applications via a Unix domain socket.
Applications register their process ID and NCCL communicator information at startup; the agent uses this to correlate eBPF-collected stacks with application-level metadata (job ID, rank, communication group).
This design avoids requiring any code changes to training scripts; only an environment variable (\texttt{SYSOM\_SOCK\_PATH}) needs to be set.

\textbf{DWARF pre-processing.}
eBPF programs run with a 512-byte stack limit and no dynamic memory allocation, precluding full DWARF CFI interpretation.
\sys{} addresses this via a two-phase approach.
At agent startup (\emph{Phase~1}), the userspace daemon parses each loaded binary's \texttt{.eh\_frame} section and extracts, for each Frame Description Entry (FDE), the CFA rule (register + offset), the return-address offset, and the PC range.
These are compiled into a sorted array loaded into an eBPF map.
FDEs using DWARF expressions (rather than simple register+offset rules) are flagged as ``complex'' and unwound via a userspace fallback; the vast majority of FDEs in our production binaries use simple rules.
At runtime (\emph{Phase~2}), the eBPF unwinder performs a binary search on the FDE array ($\lceil\log_2 M\rceil$ iterations for $M$ entries; typically 16 for $M \approx 50{,}000$), computes the CFA and return address with one memory dereference, well within eBPF limits.
Pre-processing adds $\sim$200\,ms per binary at startup.
For dynamically loaded libraries (\texttt{dlopen}), the agent detects new mappings via \texttt{/proc/[pid]/maps} polling (every 5\,s) and incrementally pre-processes the new library's \texttt{.eh\_frame} section.
JIT-compiled code (e.g., from \texttt{torch.compile}) is detected via \texttt{perf\_event\_mmap} notifications; JIT'd functions are conservatively marked as \texttt{dwarf} because their frame layout may not follow the standard ABI.
Multiple CPU cores may simultaneously encounter the same unmarked function; we use atomic compare-and-swap on the marker map so concurrent races converge to the same marker value.

\textbf{Multi-runtime stack stitching.}
AI training stacks span Python and native C++ frames.
For CPython processes, the agent locates \texttt{PyThread\-State} via the \texttt{\_PyRuntime} global symbol and thread-local storage offset, then walks the frame chain (\texttt{PyFrame\-Object.f\_back} for Python~3.10 and earlier; \texttt{\_Py\-Interpreter\-Frame} for~3.11+) to extract code object pointers and line numbers.
Native C++ frames use the hybrid FP/DWARF unwinder.
Both are stitched into a unified stack using the thread's native stack pointer as the join point.
This requires no modifications to the Python interpreter and works on unmodified production deployments.

\textbf{Data pipeline and symbol management.}
Build IDs are extracted from the ELF \texttt{.note\allowbreak.gnu\allowbreak.build-id} section.
Debug symbols are uploaded using chunked transfer (64\,MB segments) to bound peak memory on the node.
The repository uses a compact binary format with header-indexed offset and symbol data sections, supporting $O(\log n)$ lookup without loading the entire file.
Data is uploaded to a centralized log service (SLS) in batches every 30\,s.
The central service processes ingested data within minutes, computing per-group aggregates and running straggler detection.
The per-node agent has a resident memory footprint of $\sim$200\,MB, including eBPF maps and pre-processed DWARF tables.
Current deployment covers 80,000+ GPU cards across 10,000+ nodes in multiple clusters, with aggregate daily data volume of $\sim$400\,TiB.
The NCCL tracing is framework-agnostic and has been validated on Megatron (357 daily jobs), DeepSpeed (51), and ms-swift (113) in production.

\section{Evaluation}
\label{sec:evaluation}

\begin{figure}[t]
\centering
\includegraphics[width=0.85\columnwidth]{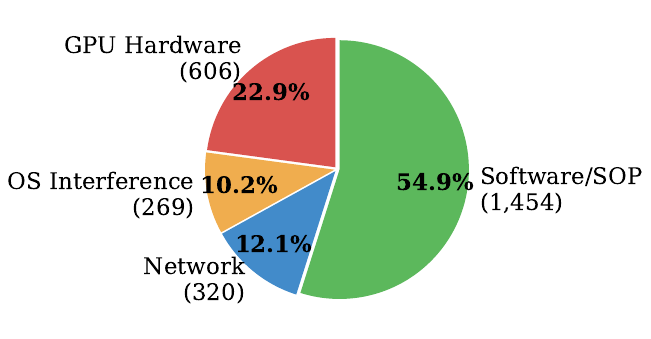}
\caption{Distribution of 2,649 diagnostic events by root-cause category during the six-month evaluation period.}
\label{fig:deployment-events}
\end{figure}

\sys{} has been deployed for over one year across 80,000+ GPUs and 10,000+ nodes at Alibaba, processing $\sim$400\,TiB of profiling data daily.
During the six-month evaluation period, the system triggered 2,649 diagnostic events (Figure~\ref{fig:deployment-events}): 606 GPU hardware issues, 269 OS interference issues (median diagnosis time 10\,min), 320 network issues (median 30\,min), and 1,454 software issues identified via log-based SOP rule matching (median 1\,min).
Of these, 94 were confirmed cross-layer incidents that required profiling-based root-cause analysis.

We evaluate the system along four dimensions: training overhead (\S\ref{sec:eval-overhead}), stack unwinding accuracy (\S\ref{sec:eval-accuracy}), symbol resolution (\S\ref{sec:eval-symbol}), and end-to-end case studies from production (\S\ref{sec:eval-cases}).

\subsection{Training Throughput Overhead}
\label{sec:eval-overhead}

\textbf{Setup.} We measure throughput impact on Llama-3.2-1B-Instruct training with 2$\times$ NVIDIA A100-80GB GPUs, PyTorch 2.1, NCCL 2.18, sequence length 1500, batch size 8.
We report the mean of 20 training steps (steps 21--40, after 20-step warm-up).
``Sampling rate'' denotes the fraction of hrtimer ticks (at 99\,Hz) that trigger a full stack collection.

\begin{table}[t]
\centering
\caption{Training throughput overhead at varying sampling rates. Baseline throughput: 15.11 iter/s.}
\label{tab:overhead}
\small
\begin{tabular}{lrr}
\toprule
\textbf{Sampling Rate} & \textbf{During Profiling} & \textbf{After Profiling} \\
\midrule
1\%   & $-$0.06\% & $-$0.20\% \\
10\%  & $-$0.33\% & $-$0.20\% \\
20\%  & $-$0.33\% & $-$0.26\% \\
40\%  & $-$0.72\% & $\phantom{-}$0\% \\
80\%  & $-$1.25\% & $-$0.20\% \\
100\% & $-$1.72\% & $-$0.26\% \\
\bottomrule
\end{tabular}
\end{table}

Table~\ref{tab:overhead} shows results.
At the production default (10\% sampling rate, equivalent to $\sim$10 samples/s), overhead is 0.33\%.
Even at 100\% (99 samples/s), overhead is only 1.72\%.
This low overhead is enabled by three factors: (1)~in-kernel stack aggregation via BPF maps eliminates per-sample userspace transitions, (2)~the hybrid unwinder avoids DWARF cost for the majority of frames, and (3)~symbol resolution is deferred to the central service rather than performed on-node.

After profiling stops, throughput returns to baseline (0--0.26\% deviation, within the 0.33\% baseline measurement noise observed with 0\% sampling).

%

\subsection{Stack Unwinding Accuracy}
\label{sec:eval-accuracy}

\begin{figure}[t]
\centering
\includegraphics[width=\columnwidth]{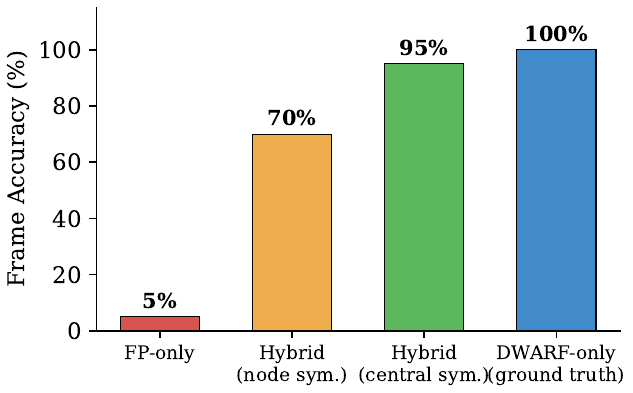}
\caption{Stack unwinding frame accuracy on production AI workloads. FP-only achieves only 5\% due to widespread omission of frame pointers. \sys{}'s hybrid unwinding with centralized symbol resolution achieves 95\%.}
\label{fig:frame-accuracy}
\end{figure}

Figure~\ref{fig:frame-accuracy} compares frame accuracy across unwinding configurations on production AI workloads.
FP-only unwinding achieves only $\sim$5\% frame accuracy because the majority of production binaries (Python, C++ with \texttt{-O2}) omit frame pointers; only Go binaries consistently preserve them.
\sys{}'s hybrid unwinding raises accuracy to 70\%, and adding centralized symbol resolution (which eliminates misattributions from sparse node-side symbol tables) further improves it to 95\%.

\subsection{Symbol Resolution}
\label{sec:eval-symbol}

\begin{figure}[t]
\centering
\includegraphics[width=\columnwidth]{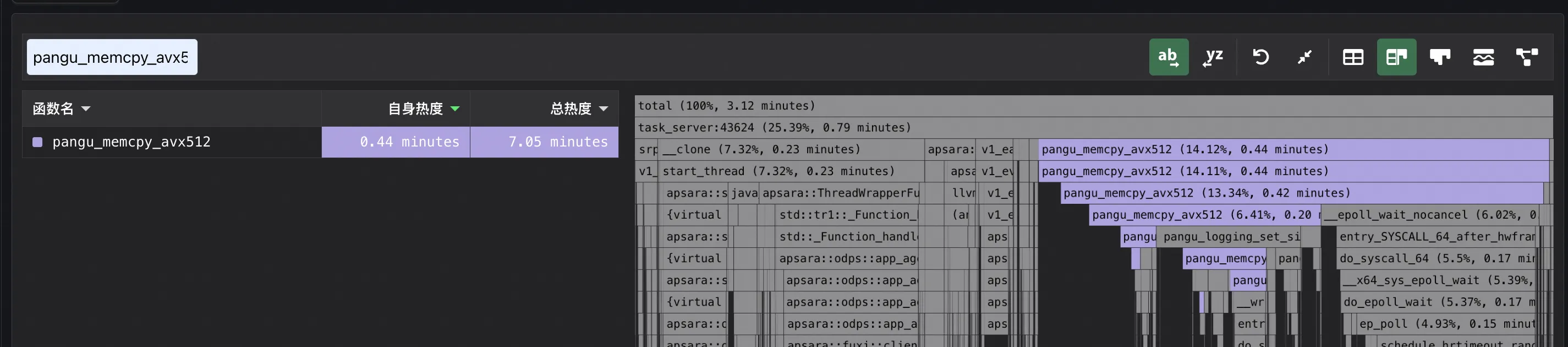}
\caption{Symbol misattribution from node-side resolution. The function \texttt{pangu\_memcpy\_avx512} absorbs samples from many unrelated functions due to sparse symbol entries, producing a fictitious hot spot. Centralized resolution with the full symbol table correctly attributes these addresses.}
\label{fig:symbol-misattrib}
\end{figure}

\textbf{Symbol misattribution case.}
In a production distributed storage workload, node-side symbol resolution caused the function \texttt{pangu\_memcpy\_avx512} (address \texttt{0x11c6aa0}, covering an 18\,MB address range due to sparse symbol entries) to absorb over 50\% of CPU samples, producing a fictitious recursive flame graph (Figure~\ref{fig:symbol-misattrib}).
With centralized resolution using the full symbol table (keyed by Build ID \texttt{ba6bc0e5...}), addresses in the \texttt{0x23XXXXXX} range were correctly attributed to dozens of distinct functions (\texttt{PrepareWatcher::Start}, \texttt{IoWatcher::onReady}, etc.), and the recursive artifact disappeared entirely.

\subsection{End-to-End Case Studies}
\label{sec:eval-cases}

We present five representative cases from production, selected to illustrate how \sys{} isolates root causes at different system layers.

\subsubsection{Case 1: GPU Thermal Throttling}

\begin{figure}[t]
\centering
\includegraphics[width=\columnwidth]{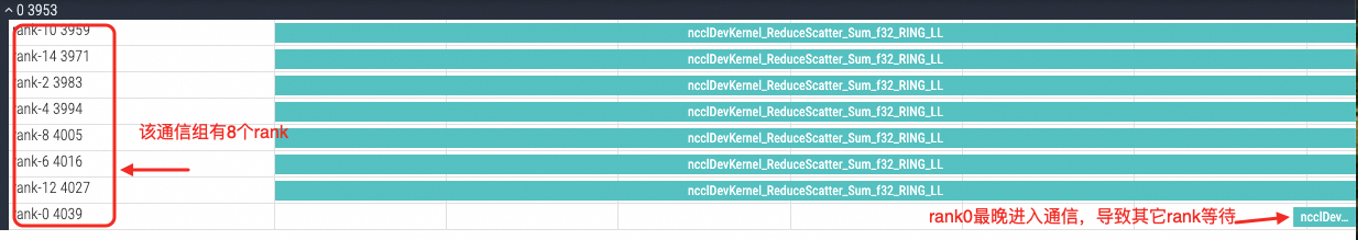}
\caption{Straggler detection for Case~1. Per-rank NCCL collective entry times show rank~0 entering last, indicating it is the straggler in this 8-rank communication group.}
\label{fig:straggler-timing}
\end{figure}

\textbf{Symptom.} Slow-node detection flagged rank~0 in a communication group of 8 ranks (ranks~0--7). Rank~0 consistently entered the \texttt{nccl\-Dev\-Kernel\_\allowbreak Reduce\-Scatter} collective 0.4\,ms later than other ranks (Figure~\ref{fig:straggler-timing}).

\begin{figure}[t]
\centering
\includegraphics[width=\columnwidth]{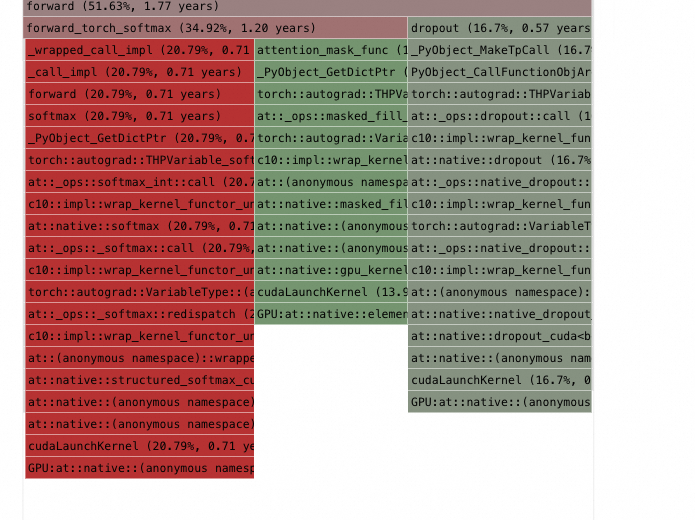}
\caption{Differential GPU flame graph for Case~1. Red frames (straggler rank~0) are uniformly wider than green frames (healthy rank~7) across all kernel types, indicating a global GPU frequency reduction rather than a specific operator issue.}
\label{fig:gpu-diff}
\end{figure}

\textbf{Diagnosis.} \sys{} generated differential GPU flame graphs between rank~0 (straggler) and rank~7 (healthy), shown in Figure~\ref{fig:gpu-diff}.
The \texttt{forward} pass on rank~0 consumed 51.6\% of GPU time (1.77\,s) vs.\ 48.1\% (1.65\,s) on rank~7.
All kernel types showed proportional slowdowns (softmax: 20.8\% vs.\ 19.1\%; dropout: 16.7\% vs.\ 15.3\%), consistent with a global frequency reduction rather than a specific operator issue.

\textbf{Root cause.} NVIDIA DCGM confirmed GPU~0 was throttled to 1,200\,MHz (vs.\ 1,410\,MHz rated) due to an ambient temperature exceedance.
Standard utilization metrics (\texttt{nvidia-smi}) reported 100\% on all ranks, masking the frequency reduction; cross-rank differential diagnosis (\S\ref{sec:differential-diagnosis}) revealed the uniform slowdown pattern, directing operators to DCGM.
After resolving the thermal issue, training throughput improved by 30\%.

\subsubsection{Case 2: NIC Soft-Interrupt Contention}

We revisit the scenario from \S\ref{sec:motivating-example} with \sys{}'s full diagnostic data.

\begin{figure}[t]
\centering
\includegraphics[width=\columnwidth]{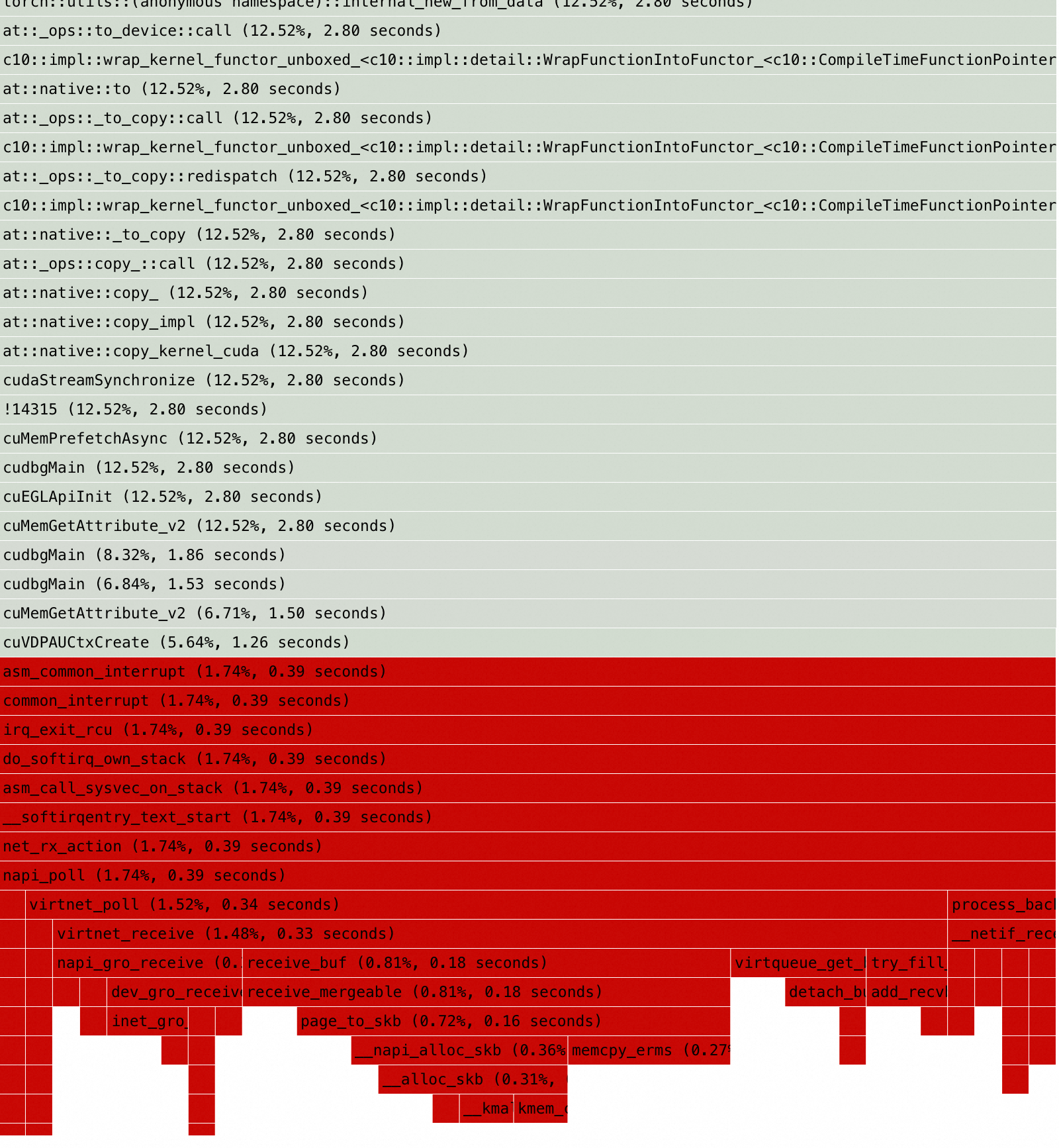}
\caption{Differential CPU flame graph for Case~2. The bottom section (red) shows the \texttt{net\_rx\_action} $\rightarrow$ \texttt{napi\_poll} $\rightarrow$ \texttt{virtnet\_receive} interrupt chain present on the straggler rank but absent on the healthy rank, identifying NIC soft-interrupt contention as the root cause.}
\label{fig:cpu-diff}
\end{figure}

\begin{sloppypar}
\textbf{Diagnosis.}
GPU diff: \sys{} compared rank~4 (straggler) and rank~6 (healthy). \emph{No significant difference}; GPU kernel times matched within 1\%.
CPU diff: Significant divergence (Figure~\ref{fig:cpu-diff}). Rank~4's CPU profile showed 1.74\% of total CPU time in the \texttt{net\_rx\_action} $\rightarrow$ \texttt{napi\_poll} $\rightarrow$ \texttt{virtnet\_poll} $\rightarrow$ \texttt{virtnet\_receive} path, with the full interrupt chain visible: \texttt{asm\_common\_interrupt} $\rightarrow$ \texttt{common\_interrupt} $\rightarrow$ \texttt{irq\_exit\_rcu} $\rightarrow$ \texttt{do\_softirq} $\rightarrow$ \texttt{net\_rx\_action} (1.74\%) $\rightarrow$ \texttt{napi\_poll} (1.74\%) $\rightarrow$ \texttt{virtnet\_poll} (1.52\%) $\rightarrow$ \texttt{virtnet\_receive} (1.48\%) $\rightarrow$ \texttt{napi\_gro\_receive} (0.81\%).
Rank~6 showed $<$0.1\% in this path.
\end{sloppypar}

\textbf{Root cause.} Network soft-interrupts (\texttt{NET\_RX\_\allowbreak SOFTIRQ}) were bound to a CPU core shared with the NCCL communication thread, causing periodic preemption.
The fix: adjust \texttt{/proc/irq/*/\allowbreak smp\_affinity} to isolate NIC interrupts from training cores.
Diagnosis took approximately 10 minutes from alert to root cause.

\subsubsection{Case 3: VFS Dentry Lock Contention}

\textbf{Symptom.}
A subset of nodes in a large training job exhibited 60\% longer iteration times.
Slow-rank detection flagged the affected ranks, but GPU kernel times matched healthy ranks within 1\%.

\begin{figure}[t]
\centering
\includegraphics[width=\columnwidth]{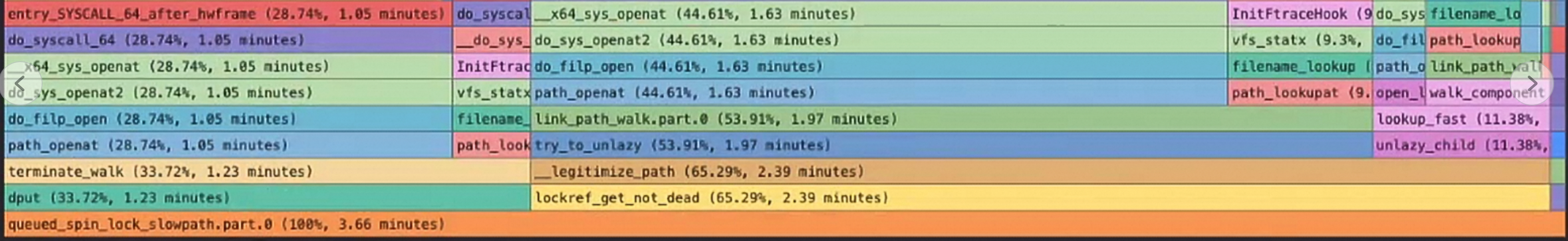}
\caption{Kernel flame graph for Case~3. All sampled CPU time is in \texttt{queued\_spin\_lock\_slowpath}, reached via VFS path lookup (\texttt{\_\_legitimize\_path}, \texttt{dput}, \texttt{lookup\_fast}). This lock contention is invisible to user-space profilers.}
\label{fig:dentry-flamegraph}
\end{figure}

\textbf{Diagnosis.}
GPU diff showed no anomaly.
CPU diff revealed that straggler ranks spent the majority of CPU time in kernel paths invisible to user-space profilers (Figure~\ref{fig:dentry-flamegraph}).
The kernel flame graph, enabled by \sys{}'s hybrid unwinding with full kernel stack support, showed 100\% of sampled time in \texttt{queued\_spin\_lock\_slowpath}, with three converging call paths:
(1)~\texttt{\_\_legitimize\_path} $\rightarrow$ \texttt{lockref\_get\_not\_dead} (65\% of samples), the RCU-to-refcount fallback triggered when dentry validation fails during lazy path walk;
(2)~\texttt{terminate\_walk} $\rightarrow$ \texttt{dput} (34\%), dentry reference release after \texttt{openat};
and (3)~\texttt{lookup\_fast} $\rightarrow$ \texttt{unlazy\_child} (11\%), child dentry lookup during path resolution.
All paths originate from \texttt{do\_sys\_openat2}, indicating system-wide VFS contention on the dentry spinlock.

\textbf{Root cause.}
A system management component periodically executed \texttt{systemctl daemon-reload} on the affected nodes, triggering massive dentry cache invalidation.
This forced all concurrent file operations (including training data loading and checkpoint I/O) to fall back from the lockless RCU path walk to the ref-counted path, causing dentry spinlock contention that blocked training threads.
This case exemplifies cross-process OS-level interference: the root cause was an unrelated system service, and diagnosing it required kernel stack visibility that no existing AI profiling tool provides (Table~\ref{tab:comparison}).

\subsubsection{Case 4: Excessive Userspace Logging Overhead}

\textbf{Symptom.}
A large-scale training job experienced a 10\% throughput drop after a routine infrastructure update.
No GPU-level anomalies were detected; all ranks showed similar iteration times (no straggler), but the absolute iteration time increased uniformly.

\textbf{Diagnosis.}
Since no straggler was flagged, \sys{} used temporal baseline comparison (\S\ref{sec:differential-diagnosis}), comparing the current per-group flame graph against the pre-update baseline.
The CPU flame graph showed a new hot path in \texttt{SLS::LogClient::Send} $\rightarrow$ \texttt{protobuf::Serialize} $\rightarrow$ \texttt{memcpy}.
This path was absent in the pre-update baseline.

\textbf{Root cause.}
The infrastructure update increased the verbosity of Alibaba's SLS (Simple Log Service) client from \texttt{INFO} to \texttt{DEBUG}, adding per-iteration logging of tensor statistics.
The serialization overhead consumed CPU cycles on the training threads, uniformly slowing all ranks.
The fix: revert the log level.
Because all ranks slowed uniformly, no straggler alert fired; always-on profiling enabled temporal baseline comparison (\S\ref{sec:differential-diagnosis}) to surface the new hot path.

\subsubsection{Case 5: Data Ingestion Bottleneck}

\textbf{Symptom.}
A large-scale training job reported a 30\% throughput drop with no changes to model configuration or cluster topology.

\textbf{Diagnosis.}
NCCL observability (\S\ref{sec:nccl-observability}) confirmed that collective latencies were uniform across ranks, ruling out communication and straggler issues.
Instead, the CPU flame graph, with accurate stacks from hybrid unwinding (\S\ref{sec:hybrid-unwinding}) and centralized symbol resolution (\S\ref{sec:symbol-resolution}), revealed elevated time in \texttt{cpfs} (a cloud-native parallel file system client) and \texttt{ossutils} (object storage utilities), far above the normal baseline.

\textbf{Root cause.}
The data pipeline's throughput had degraded due to increased dataset size without a corresponding storage-tier upgrade.
I/O-bound data loading saturated host CPU resources, uniformly slowing all ranks.
The fix: upgrade the storage backend and increase data-loader parallelism.
GPU profiling alone would have shown normal kernel times. Attributing the slowdown to the data pipeline required the combined view of normal NCCL timing and elevated host-side CPU in storage I/O paths (\S\ref{sec:differential-diagnosis}).
We have observed similar I/O-related slowdowns in multiple other incidents: in one case, object storage (OSS) rate limiting caused by co-tenant contention reduced training throughput by 30\%; in another, slow network fetches of training data caused a 9$\times$ throughput drop.

\section{Related Work}
\label{sec:related}

\textbf{AI training and GPU profiling.}
Table~\ref{tab:comparison} compares \sys{} with existing tools.
MegaScan~\cite{megascan} traces Python-level Megatron execution; MegaScale~\cite{jiang2024megascale} collects GPU kernel runtimes; Nsight Systems~\cite{nsight} provides GPU timelines at 10--30\% overhead~\cite{nsight_overhead}; Strobelight~\cite{strobelight} profiles C++ stacks without GPU or Python visibility; DeepContext~\cite{zhao2025deepcontext} links Python contexts to GPU code but lacks OS kernel visibility; Flare~\cite{cui2025flare} deploys full-stack tracing across 6,000 GPUs but focuses on fault detection rather than continuous profiling.
All operate at the process level without kernel stack visibility.
For GPU-internal analysis, Neutrino~\cite{huang2025neutrino} and KPerfIR~\cite{guan2025kperfir} provide instruction-level GPU kernel profiling and are complementary to \sys{}'s host-side approach.
Recent characterization studies~\cite{hu2024characterization,patel2024revisiting} catalog training failures but do not provide a diagnostic system.

\textbf{Continuous profiling and distributed tracing.}
Google-Wide Profiling~\cite{gwp2010} pioneered always-on fleet profiling; Parca~\cite{parca} and Pyroscope~\cite{pyroscope} provide eBPF-based continuous profiling but use FP-only unwinding, limiting accuracy.
Elastic Universal Profiling~\cite{elastic_profiling} pre-processes DWARF for eBPF but does not provide per-function adaptive selection.
Meta's Dynolog~\cite{dynolog2022} integrates CPU/GPU telemetry but does not address unwinding accuracy.
The Linux kernel's ORC unwinder~\cite{orc} pre-computes kernel-space unwind tables; FP unwinding~\cite{gregg2020bpf} is $O(1)$ per frame, whereas DWARF~\cite{dwarf5} is accurate but expensive; \sys{}'s contribution is the adaptive per-function combination.
Causal profiling (Coz~\cite{curtsinger2015coz}, BCOZ~\cite{ahn2024bcoz}) and wait-for analysis (wPerf~\cite{zhou2018wperf}) target single-node bottlenecks; \sys{} addresses the complementary problem of cross-layer, cross-rank diagnosis.
In distributed tracing, Fay~\cite{erlingsson2011fay} established cross-layer tracing from kernels to clusters; Pivot Tracing~\cite{mace2015pivot}, lprof~\cite{zhao2014lprof}, Stitch~\cite{zhao2016stitch}, and Hubble~\cite{luo2022hubble} reconstruct distributed request flows; EXIST~\cite{wang2025exist} provides low-overhead intra-service tracing; Jaeger~\cite{jaeger}, Zipkin~\cite{zipkin}, and Sage~\cite{sage2021} correlate spans in microservice settings.
These target request-oriented systems; \sys{} targets bulk-synchronous GPU training with per-rank profiling rather than per-request tracing.

\section{Limitations}
\label{sec:discussion}


\sys{} relies on Linux eBPF (kernel 5.4+), CUDA runtime uprobes, and NCCL structure layouts (currently covering NCCL~2.14--2.21 and ACCL).
Python frame extraction supports CPython~3.10+; JIT runtimes that do not emit standard ELF mappings are not supported.
Straggler detection assumes a small number of anomalous ranks per communication group; when a majority degrade simultaneously, the statistical outlier model loses power and temporal baseline comparison is needed.
Diagnostic coverage varies by root-cause type: GPU-side stalls are reliably detected; on-CPU interference is covered by flame graph analysis; however, off-CPU blocking (e.g., a training process waiting on slow data reads) is not captured by on-CPU sampling, and RDMA-level network issues can be identified as slow collectives but not root-caused without additional network-layer instrumentation.
\sys{} captures GPU kernel timing via host-side CUDA events but not GPU-side call stacks; finer-grained analysis requires complementary tools such as Nsight Compute~\cite{nsight}.
The central analysis service is a single cluster without geographic redundancy; if unavailable, node agents buffer data locally for up to 1 hour.

\section{Conclusion}
\label{sec:conclusion}

OS-level stalls (scheduler contention, soft-interrupts, memory pressure) are a substantial source of AI training performance issues, yet existing tools lack continuous OS-kernel visibility correlated with GPU and communication activity.
We presented \sys{}, a production cross-layer observability system that addresses this gap through adaptive hybrid stack unwinding, centralized symbol resolution, framework-agnostic NCCL observability, and layered differential diagnosis.
Deployed across 80,000+ GPUs at Alibaba for over one year, \sys{} achieves less than 0.4\% overhead, 95\% frame accuracy, and has helped diagnose 94 confirmed production issues, reducing median diagnosis time from days to approximately 10 minutes.
We believe the cross-layer, always-on profiling methodology generalizes beyond AI training to other large-scale distributed systems where OS-level interference is difficult to isolate.

\bibliographystyle{ACM-Reference-Format}
\bibliography{references}

\end{document}